%% file: ms.tex
\documentclass[twocolumn, floatfix]{aastex701}
\usepackage{amsmath}
\usepackage{amssymb}
\usepackage{amsfonts}
\usepackage{graphicx}
\usepackage{multirow}
\usepackage{booktabs}
\usepackage{subcaption}

\begin{document}
\shortauthors{The SPOTLIGHT Collaboration}
\title{Real-Time RFI Mitigation in SPOTLIGHT: A Two-Stage Approach for Transient Searches}
\shorttitle{The SPOTLIGHT's Real-time RFI Mitigation Framework}

\correspondingauthor{Raghav Wani}
\affiliation{National Centre for Radio Astrophysics (NCRA), Pune-411007, Maharashtra, India}
\affiliation{Indian Institute of Science Education and Research (IISER), Pune-411008, Maharashtra, India}
\email{raghav.wani@gmail.com, wani.raghav@students.iiserpune.ac.in}

\author[0009-0002-2515-2425]{Raghav Wani}
\affiliation{National Centre for Radio Astrophysics (NCRA), Pune-411007, Maharashtra, India}
\affiliation{Indian Institute of Science Education and Research (IISER), Pune-411008, Maharashtra, India}
\email{raghav.wani@gmail.com, wani.raghav@students.iiserpune.ac.in}

\author[0000-0002-2892-8025]{Jayanta Roy}
\email{jroy@ncra.tifr.res.in}
\affiliation{National Centre for Radio Astrophysics (NCRA), Pune-411007, Maharashtra, India}

\author[0000-0002-7551-5215]{Harshavardhan Reddy}
\affiliation{National Centre for Radio Astrophysics (NCRA), Pune-411007, Maharashtra, India}
\email{reddysh@gmrt.ncra.tifr.res.in}

\author[0000-0002-6631-1077]{Sanjay Kudale}
\affiliation{National Centre for Radio Astrophysics (NCRA), Pune-411007, Maharashtra, India}
\email{kudale.sanjay@gmail.com}

\author[0000-0002-2441-4174]{Ujjwal Panda}
\email{ujjwalpanda97@gmail.com, upanda@ncra.tifr.res.in}
\affiliation{National Centre for Radio Astrophysics (NCRA), Pune-411007, Maharashtra, India}

\author[0000-0003-2797-0595]{Karel Adamek}
\affiliation{Department of Physics, Silesian University in Opava, Opava, 74601, Czech Republic}
\email{karel.adamek@gmail.com}

\author[0000-0003-1756-3064]{Wesley Armour}
\affiliation{Oxford e-Research Centre (OeRC), University of Oxford, Oxford-OX13PJ, United Kingdom}
\email{wes.armour@oerc.ox.ac.uk}

\author[0000-0002-8550-9070]{Kshitij Bane}
\affiliation{National Centre for Radio Astrophysics (NCRA), Pune-411007, Maharashtra, India}
\email{kshitijbane@gmail.com}

\author[0000-0003-3445-4521]{Kaushal Buch}
\affiliation{National Centre for Radio Astrophysics (NCRA), Pune-411007, Maharashtra, India}
\email{kdbuch@gmrt.ncra.tifr.res.in}

\author[0000-0002-0269-1154]{Jayaram Chengalur}
\affiliation{National Centre for Radio Astrophysics (NCRA), Pune-411007, Maharashtra, India}
\email{chengalur@ncra.tifr.res.in}

\author[0009-0006-7995-5871]{Jyotirmoy Das}
\affiliation{National Centre for Radio Astrophysics (NCRA), Pune-411007, Maharashtra, India}
\email{tataidas5392@gmail.com}

\author[0000-0003-3747-9847]{Sridhar Gajendran}
\affiliation{National Centre for Radio Astrophysics (NCRA), Pune-411007, Maharashtra, India}
\email{sridhar.gajendran@gmail.com}

\author[]{SPOTLIGHT Collaboration}
\email{spotlight@ncra.tifr.res.in}
\affiliation{National Centre for Radio Astrophysics (NCRA), Pune-411007, Maharashtra, India}

\input{sections/abstract.tex}
\input{sections/intro.tex}
\input{sections/strategy.tex}
\input{sections/results.tex}
\input{sections/futurework.tex}
\input{sections/summary.tex}
\input{sections/thanks.tex}
\input{sections/appendix.tex}

\bibliography{refs}{}
\bibliographystyle{aasjournal}

\end{document}

%% file: sections/abstract.tex
\begin{abstract}
Radio Frequency Interference (RFI) remains one of the primary challenges limiting the sensitivity and reliability of modern radio transient surveys, particularly for real-time searches of fast radio transients. The SPOTLIGHT system is a commensal real-time transient search backend operating at the upgraded Giant Metrewave Radio Telescope (uGMRT), where robust and computationally efficient RFI mitigation is essential for sustained operations. We present the real-time two-stage RFI mitigation framework developed for SPOTLIGHT, comprising an antenna-level voltage-filtering module (\texttt{VOLT}) operating prior to correlation beamforming and the SPOTLIGHT Time-domain RFI Processing Engine (\texttt{STRIPE}), a statistical RFI-filtering framework applied to beamformed data. Together, these complementary techniques mitigate a broad spectrum of RFI, ranging from broadband impulsive interference mitigated by \texttt{VOLT} to narrowband spectrally confined spurious signals mitigated by \texttt{STRIPE}, while remaining computationally efficient enough to satisfy the stringent requirements of real-time processing. The framework is evaluated using routine commensal GMRT observations, controlled $76$ pulsar observations, and benchmarking against PRESTO's \texttt{rfifind}. The deployed system reduced the false detection rate by 98\%. The recovered astrophysical pulses exhibit a 2.7$\times$ improvement in S/N after the two-stage filtering compared with the unfiltered data. These improvements enhance SPOTLIGHT's detection efficiency, sensitivity, and operational reliability, strengthening its capability to discover radio transients with the uGMRT.
\end{abstract}

%% file: sections/intro.tex
\section{Introduction}

Radio telescopes are, inadvertently, affected by human-made sources of interference. As industries, power grids, telecommunication systems, and digital infrastructure have expanded, the amount of human-made radio frequency interference (RFI) has increased drastically, including the emergence of entirely new sources of RFI (for instance, wind turbines \citep{brentjens_interference_2016, winkel_compatibility_2019}, lightning \citep{sokolowski_statistics_2015}, and satellite constellations \citep{vruno_unintended_2023, bassa_bright_2024}). Thus, there has been a continuous investment for the development of more and better RFI mitigation methodologies, resulting in a wide range of algorithms \citep{leshem_multichannel_2000, raza_spatial_2002, fridman_rfi_2001, bentum_rfi_2016, offringa_post-correlation_2010, baan_implementing_2019, cucho-padin_radio_2019, akeret_radio_2017, vos_generative_2019}. A majority of these algorithms are offline; that is, they operate on data stored on disk. This offers an advantage, since the full-range of the data in frequency and time can be accessed, and the data can be accessed multiple times, and multiple passes can be made through the data, if necessary. This has allowed the development of sophisticated algorithms for offline data \citep{offringa_post-correlation_2010, maan_fourier_2021}, including machine learning (ML) and deep learning (DL) based approaches \citep{akeret_radio_2017, vos_generative_2019, mesarcik_deep_2020}. However, developing algorithms that can operate on data streams in real-time is tough, due to several factors: firstly, such algorithms can only access the data once, and the amount of data being accessed at any given instance is small. Thus, such algorithms need to successfully determine and mitigate the presence of RFI from a limited instance of data in a single pass. Such algorithms have been developed, often with the help of robust statistical estimators, such as skewness and kurtosis \citep{nita_radio_2007, gary_wideband_2010, nita_generalized_2010, nita_spectral_2016}, median absolute deviation (MAD) \citep{buch_real-time_2019}, the autocorrelation function (ACF) \citep{morello_iqrm_2021}, and others \citep{van_nieuwpoort_real-time_2018, sclocco_real-time_2019}. These implementations also need to be computationally efficient, in order to keep with real-time requirements; this has led to the development of algorithms that specifically meet this requirement \citep{gong_mars_2026}, as well as re-implementation of preexisting algorithms on hardware accelerators such as FPGAs \citep{baan_radio_2004, buch_real-time_2019} and GPUs \citep{van_nieuwpoort_towards_2016, van_nieuwpoort_real-time_2018, sclocco_real-time_2019} for faster performance.

In this work, we present exactly such a real-time RFI mitigation framework, developed for the SPOTLIGHT project\footnote{\url{https://spotlight.ncra.tifr.res.in}} \citep{roy_spotlight_2024}, a commensal survey for radio transients being undertaken at the upgraded Giant Metrewave Radio Telescope (uGMRT; \citep{gupta_upgraded_2017}). This survey is carried out using a dedicated high-performance cluster (HPC) comprising 60 Rudra servers and 90 NVIDIA A100 GPUs. 640 post-correlation (PC) phased array beams \citep{roy_post-correlation_2018} are formed by SPOTLIGHT's correlator/beamformer (Reddy et al. (in prep.)), optimally tiled across the field-of-view; alongside, a single incoherent array (IA) beam is formed as well. Both the PC beams and the IA beam are searched for transients in real-time using a highly-optimised search pipeline \citep{panda_spotlight_2026}; 160 out of 640 beams are recorded to disk, and are put through an offline pulsar search pipeline \citep{das_spotlight_2026}.

The correlator/beamformer and the transient search system each use 32 GPUs, while the pulsar search pipeline uses 10 GPUs; the remaining GPUs are used by other SPOTLIGHT modules currently in development. Thus, it became necessary to develop a RFI mitigation framework that would meet the required computational requirements, without relying on GPUs. This framework was developed in two stages: the first stage operates on Nyquist-sampled voltage time series from individual antennas, while the second stage performs RFI mitigation on the beamformed data products. A detailed description of this two-stage strategy can be found in \S\ref{sec:two_stage}. Then, results from our benchmarks using real observational datasets obtained with the SPOTLIGHT system are discussed in \S\ref{sec:results}. Finally, our plans for future improvements are detailed in \S\ref{sec:futurework}, and our conclusions are summarised in \S\ref{sec:summary}.

%% file: sections/strategy.tex
\section{SPOTLIGHT's RFI filtering strategy}\label{sec:two_stage}

\begin{figure}
    \centering
    \includegraphics[width=\linewidth]{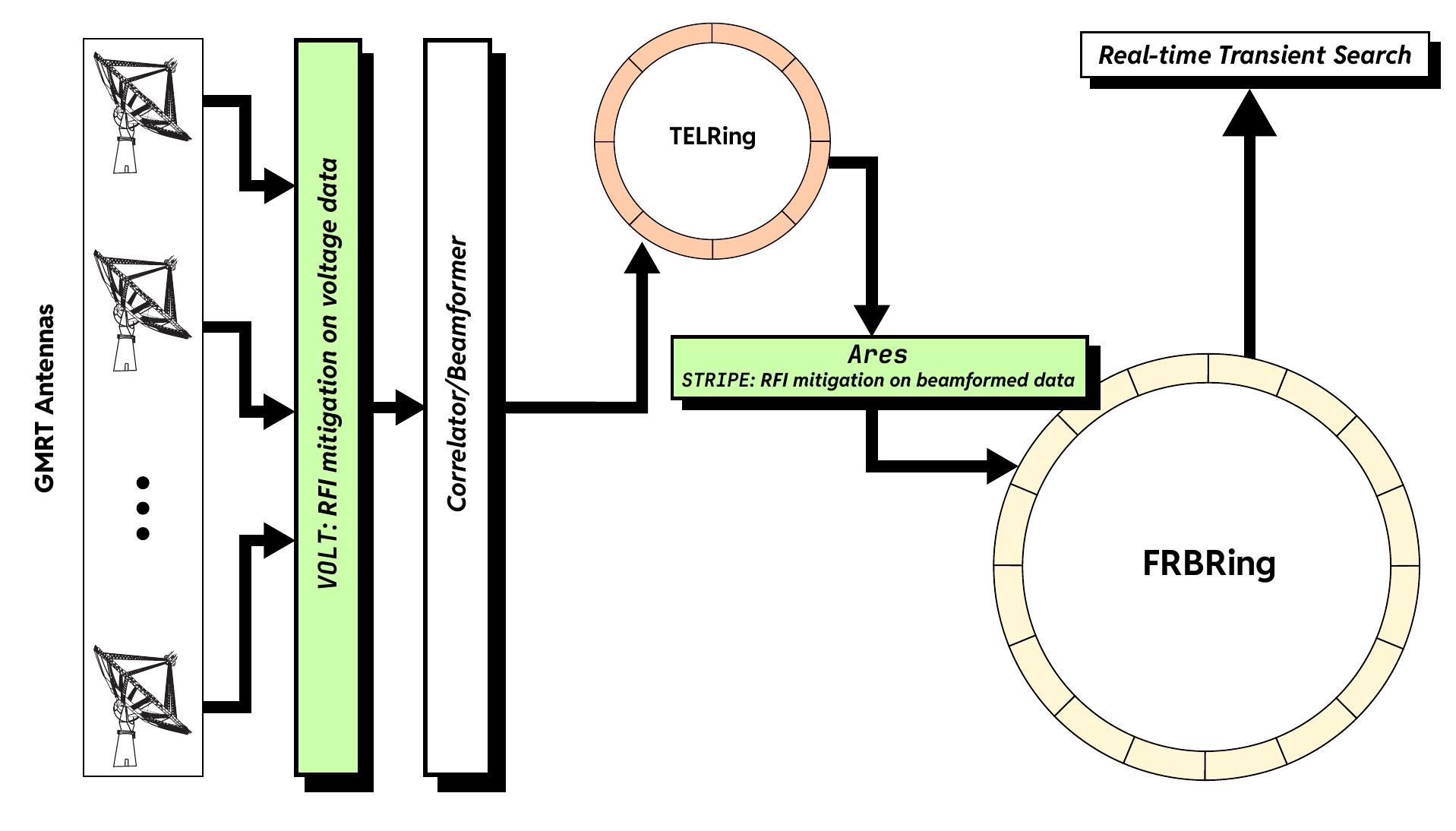}
    \caption{Signal-processing flow of the SPOTLIGHT backend, showing the locations of the two-stage RFI mitigation framework. \texttt{VOLT} operates on antenna voltages prior to correlation and beamforming, while \texttt{STRIPE} mitigates residual RFI in the beamformed data during transfer from \texttt{TELRing} to \texttt{FRBRing}, before real-time transient searching.}
    \label{fig:flowchart}
\end{figure}

SPOTLIGHT employs a two-stage real-time RFI mitigation framework that suppresses interference both before and after beamforming, enabling efficient removal of complementary classes of RFI while satisfying the stringent latency requirements of commensal transient searches. Broadband impulsive interference is first mitigated at the antenna-voltage level at Nyquist resolution prior to correlation, whereas residual broadband contamination, persistent narrowband interference, and slowly varying baseline fluctuations are removed from the beamformed data before transient searching.

In the first stage, digitised antenna voltages are processed by the voltage-level RFI filter, referred to hereafter as \texttt{VOLT}, adapted from the real-time broadband RFI excision framework developed for the GMRT Wideband Backend (GWB, \cite{reddy_wideband_2017}) in \cite{buch_real-time_2019} and \cite{buch_real-time_2023}. \texttt{VOLT} is a multi-CPU implementation of the original FPGA-based excision module developed for GWB \citep{buch_real-time_2019}. Since \texttt{VOLT} operates directly on Nyquist-sampled voltages before cross-correlation, it efficiently suppresses strong impulsive broadband interference that would otherwise propagate coherently through the beamformer. The filtered antenna voltages are subsequently correlated within the SPOTLIGHT correlator and beamformer to generate interferometric visibilities, which are then used to form 640 simultaneous high-sensitivity post-correlation tied-array beams. The beamformed data are then processed by the SPOTLIGHT Time-domain RFI Processing Engine (\texttt{STRIPE}) prior to transient searching. Unlike \texttt{VOLT}, \texttt{STRIPE} operates on beamformed dynamic spectra and employs statistical techniques to identify and suppress residual broadband and narrowband interference that remains after beamforming. 

Figure \ref{fig:flowchart} illustrates the position of \texttt{VOLT} and \texttt{STRIPE} within the SPOTLIGHT system. The shared-memory architecture of SPOTLIGHT has been described in detail by \cite{panda_spotlight_2026} and is summarised here only briefly. Following first-stage antenna voltage RFI filtering by \texttt{VOLT}, correlation and beamformation, the beamformed data are initially written to a smaller shared-memory ring buffer, referred to as \texttt{TELRing}, from where they are transferred by the program \texttt{Ares}\footnote{\url{https://github.com/nsmspotlight/Ares}} into a larger shared-memory buffer, \texttt{FRBRing}. During this transfer, each block of beamformed data is sequentially scaled and processed by \texttt{STRIPE}, along with other observation-specific operations as outlined in \cite{panda_spotlight_2026}, and finally written into \texttt{FRBRing}. The real-time transient search pipeline subsequently operates exclusively on the RFI-mitigated data stored in this larger buffer \citep{panda_spotlight_2026}.

\texttt{TELRing} consists of eight buffer blocks that temporarily store 8.388608 seconds of beamformed data. In contrast, \texttt{FRBRing} stores twelve larger blocks, with each block equivalent to four complete \texttt{TELRing} buffers, and thus stores 8.388608 $\times$ 12 $\times$ 4 $=$ 402.653184 seconds of observations. Such a large buffer accommodates the latency for downstream transient detection, candidate validation, and simultaneous dumping of beamformed data, visibilities, and baseband voltages once a candidate has been identified. This architecture enables continuous real-time processing without interrupting the incoming data stream while accommodating the latency associated with downstream transient search and validation.

The complementary operations of \texttt{VOLT} and \texttt{STRIPE} provide a real-time RFI mitigation framework for SPOTLIGHT, improving data quality while preserving the computational efficiency required for continuous commensal observations. Detailed descriptions of the two modules are presented in the following subsections.

\input{sections/voltintro.tex}
\input{sections/stripeintro.tex}

%% file: sections/voltintro.tex
\subsection{\texttt{VOLT}: RFI filtering on voltage data}\label{sec:volt}

In order to improve robustness against long-duration RFI, \texttt{VOLT} employs the Median-of-MAD (MoM) estimator proposed in \cite{buch_implementing_2019} and in \cite{buch_real-time_2023}, wherein the thresholds for mitigation are determined from the median of median absolute deviation (MAD) estimates obtained from successive data windows, rather than from a single window. It has been shown in \cite{buch_implementing_2019} that this estimator remains robust even when $\approx 50\%$ of the data is contaminated by RFI. For an input block of digitised voltage samples, $x$, the median $M(x)$ and the MAD, $D(x)$ are estimated. Then, the upper and lower thresholds are computed as:
\begin{align*}
    \tau_{U} &= M(x) + n \times 1.4826 D(x), \\
    \tau_{L} &= M(x) - n \times 1.4826 D(x),
\end{align*}
where $n$ represents a user-defined threshold multiplier. The MAD $D(x)$ is scaled by a factor of 1.4826 to yield the equivalent standard deviation, assuming that the data follows a normal distribution. Under the same assumption, the median $M(x) \approx 0$, and hence its calculation can be avoided entirely. Then, the MoM estimator, $M_{D}$, is calculated across $k$ successive windows, as:
\begin{equation*}
    M_{D} = M(D_{1}, D_{2}, \dots, D_{k})
\end{equation*}
where $D_{1}, D_{2}, \dots, D_{k}$ are the MAD values for each of the $k$ windows. In the SPOTLIGHT system, for a bandwidth of 200 MHz and a corresponding Nyquist resolution of 2.5 nanoseconds, the MoM estimator is calculated over 20,480 MAD values, and each MAD value is calculated over 20,480 voltage samples. For 100 MHz, the number of MAD values used remains the same, but the number of samples used to calculate each MAD value becomes 10,240. In both cases, the total amount of data used to estimate the thresholds becomes $20,480 \times 20,480 \times 2.5 \text{ ns} = 20,480 \times 10,240 \times 5 \text{ ns} = 1.048576 \text{ s}$, which is exactly equal to the duration of one block of the \texttt{TELRing} buffer. Synchronising the processing of voltage data with the beamformer's block boundaries simplifies the data flow through the system and enables continuous low-latency operation throughout.

The current implementation of \texttt{VOLT} is only applicable when SPOTLIGHT observes in its 100 or 200 MHz modes. When operating at a bandwidth of 400 MHz, \texttt{VOLT} is bypassed entirely. This limitation is primarily computational: for 400 MHz, one must process 40,960 samples in a single pass, while meeting the same real-time requirements. Another restriction arises from memory availability: in order to fit 400 MHz voltage samples in memory, they are packed as 4-bit integers. However, these samples need to be unpacked into 8-bit integers during RFI excision and repacked into the 4-bit representation before being sent downstream to the correlator/beamformer. In GWB, these issues are avoided by falling back to the RFI excision algorithm's original FPGA implementation \cite{buch_implementing_2019}, which has none of these restrictions. Work on optimising \texttt{VOLT} is currently underway.

In its present implementation, \texttt{VOLT} processes dual-polarisation complex voltage streams from 32 antennas (30 real antennas + 2 dummy antennas), for a total of 64 baseband inputs. These are uniformly distributed across 16 nodes. On each node, 4 OpenMP CPU threads are employed per input. The RFI-mitigated voltages are then supplied to SPOTLIGHT's correlator/beamformer, which turns them into channelised, beamformed data.

%% file: sections/stripeintro.tex
\subsection{\texttt{STRIPE}: RFI filtering on beamformed data}\label{sec:stripe}

\begin{figure}
    \centering
    \includegraphics[width=1\linewidth]{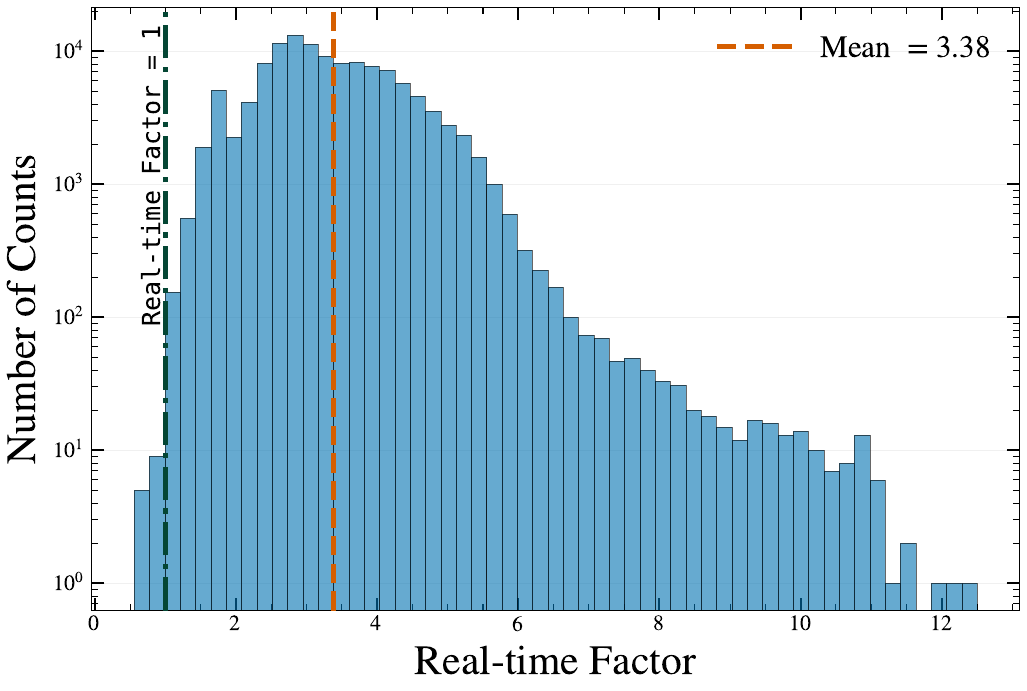}
    \caption{Distribution of the real-time speed-up factor of \texttt{STRIPE} for 640 post-correlation beams over 35.8 hours of commensal GTAC observations. Each measurement corresponds to a 1.048576 s \texttt{TELRing} block}
    \label{fig:stripe_rtf_dist}
\end{figure}

The SPOTLIGHT Time-domain RFI Processing Engine (\texttt{STRIPE})\footnote{\url{https://github.com/RaghavWani/STRIPE}} is a lightweight statistical RFI mitigation framework, developed specifically for SPOTLIGHT's beamformed data. \texttt{STRIPE} is adapted from \texttt{PulsarX}'s \texttt{filtool} utility \citep{men_pulsarx_2023}, modifying its implementation as per SPOTLIGHT's shared-memory-based ring buffer architecture, as well as its real-time constraints. It is deployed as a sub-module of the \texttt{Ares} utility, which transfers data from the smaller \texttt{TELRing} to the larger \texttt{FRBRing}, the buffer read by SPOTLIGHT's real-time transient search system; more details can be found in \cite{panda_spotlight_2026}. As a part of \texttt{Ares}, in each iteration \texttt{STRIPE} processes a single block of \texttt{TELRing} consisting of 800 samples, across 16 nodes. On each node, data from 40 PC beams is processed across 16 CPU cores; thus, a single block of data from $40 \times 16 = 640$ PC beams is processed across $16 \times 16 = 256$ CPU cores. \texttt{STRIPE} processes each \texttt{TELRing} block 3.4 times faster than real time on average, as seen in in Figure \ref{fig:stripe_rtf_dist}, which shows the distribution of real-time speed-up factors per block. The real-time speed-up factor for each run is calculated as the ratio of the duration of each block by the time taken to process each block. Although the buffer occasionally approaches exhaustion (that is, the filter performance drops below 1), the \texttt{TELRing} buffer stores 8.388608 seconds of data, so the filter can catch up, and no data is lost. \texttt{STRIPE}'s computational efficiency leaves sufficient resources available for SPOTLIGHT's transient search system downstream. For the IA beam, a single node and 16 CPU cores are used to process each block. Unlike \texttt{VOLT}, \texttt{STRIPE} works for all SPOTLIGHT configurations.

\begin{figure*}
    \centering
    \includegraphics[width=1\linewidth]{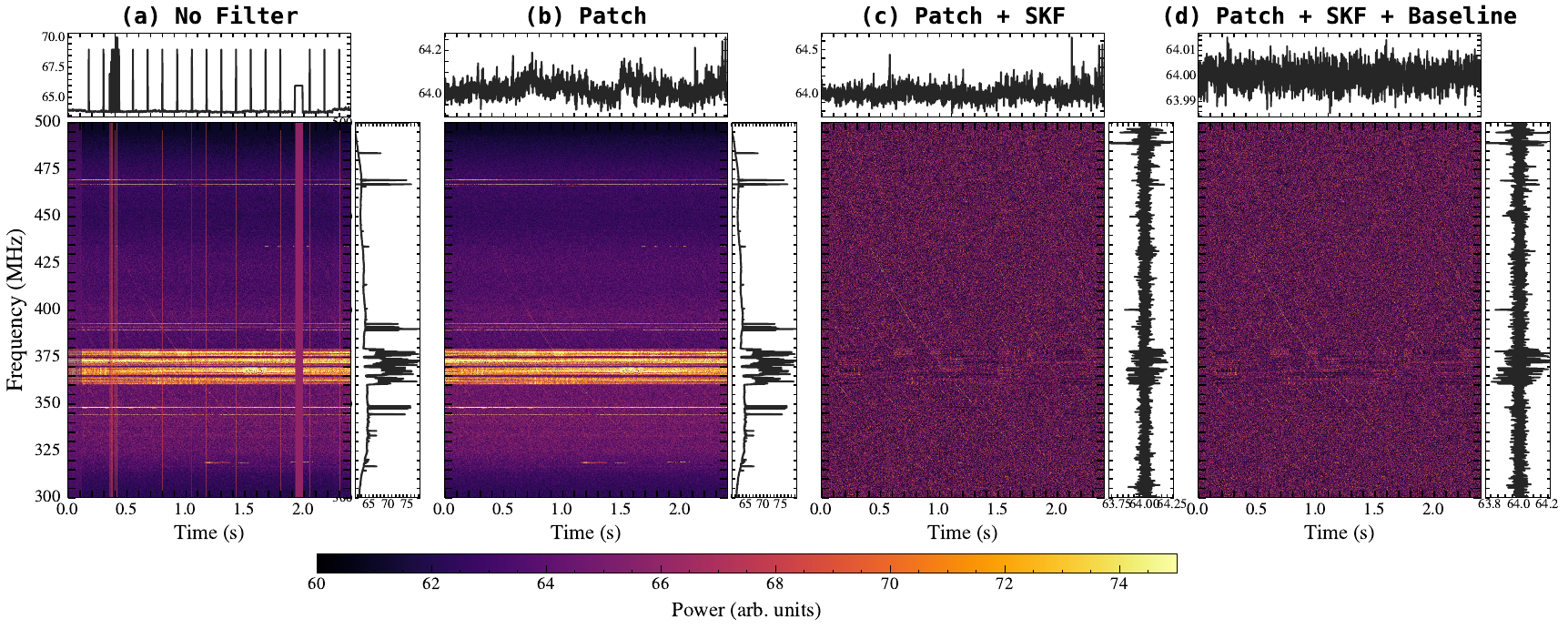}
    \caption{Step-by-step RFI mitigation in representative Band 3 (300–500 MHz) beamformed data using the three \texttt{STRIPE} filtering stages. Each panel shows the dynamic spectrum together with the corresponding time- and frequency-collapsed profiles. The successive stages progressively suppress RFI while preserving the underlying dispersed pulsar signal.}
    \label{fig:stripe_filters}
\end{figure*}

Currently, \texttt{STRIPE} adapts \texttt{filtool}'s patch, skewness-kurtosis, and baseline filters. The patch filter identifies time samples exhibiting zero variance across the entire band, indicating that these samples are contaminated by impulsive, broadband RFI, since their power is uniform across the band. The identified samples, along with their immediate neighbours, are then replaced with Gaussian noise generated according to each channel's mean and variance, estimated from unflagged samples. This helps preserve the statistical properties of the data. The effect of the patch filter can be seen in Figure \ref{fig:stripe_filters}b. The data is then passed to the skewness-kurtosis (SK) filter.

The SK filter follows the implementation in \texttt{filtool}. The filter identifies narrowband RFI by calculating the spectral skewness, the spectral excess kurtosis, and the spectral autocorrelation function with a 1-sample delay (hereafter referred to as $\mathrm{ACF}1$). All of these statistics measure deviation from ideal, uncorrelated Gaussian noise. A large value of skewness and kurtosis indicates that the distribution of the data is highly non-Gaussian; similarly, a non-zero value of the autocorrelation function indicates that there are correlations in the data, which should be ideally be absent (indeed, for uncorrelated Gaussian noise, $\mathrm{ACF} = 0$, regardless of time lag). Thus, the SK filter calculates these statistics for each frequency-time block and identifies outliers in all three using Tukey's fences \citep{tukey_exploratory_1977}. This is a robust non-parametric method for outlier detection, and is based on the interquartile range (IQR). For any arbitrary metric $M$, channel $j$ is considered to be uncontaminated if:
\begin{equation*}
    Q_{1}(M) - \kappa \times \mathrm{IQR(M)} \leq M_{j} \leq Q_{3}(M) + \kappa \times \mathrm{IQR(M)},
\end{equation*}
where $Q_{1}$ and $Q_{3}$ are the first and third quartiles, respectively, and $\mathrm{IQR} = Q_{3} - Q_{1}$. The parameter $\kappa$ is a user-defined threshold that controls the aggressiveness of the outlier rejection and is the only tunable parameter in the SK filter. A sample is only flagged if all three statistical measures satisfy the above criterion. The flagged sample is then replaced with zero-mean, unit-variance Gaussian noise, while all other samples are normalised to zero mean and unity variance, as per the spectral mean and variance obtained from the data. This ensures that, for each frequency-time block, the bandpass is normalised and the data statistics are preserved. The effect of the SK filter is shown in Figure \ref{fig:stripe_filters}c. The data is then passed to the baseline filter.

The baseline filter removes slow-varying fluctuations in time that are introduced by instrumental or other effects. It does so by first estimating the baseline using a running median filter. For each frequency-time block, $D(\nu, t)$, a one-dimensional time series, $s(t)$, is obtained by averaging across all frequency channels. This helps suppress any channel-specific fluctuations that will affect the baseline estimate. Then, a running median filter with width $\omega$ is run over this time series, and a robust estimate of the baseline, $\tilde{s}(t)$, is obtained. Note that it can be ensured that the running median filter is insensitive to short-duration impulsive events, by keeping the width large enough with respect to the typical duration of such events. Then, the baseline for each channel is modelled independently using the linear relation:
\begin{equation*}
    B_{\nu}(t) = \alpha_{\nu} \tilde{s}(t) + \beta_{\nu},
\end{equation*}
where the coefficients $\alpha_{\nu}$ and $\beta_{\nu}$ are determined through a simple linear least squares fit for each channel. The baseline is then subtracted from each frequency-time block as follows:
\begin{equation*}
    D\prime(\nu, t) = D(\nu, t) - B_{\nu}(t)
\end{equation*}
The only free parameter for the baseline filter is the width, $\omega$, used by the running median filter. The effect of the baseline filter is shown in Figure \ref{fig:stripe_filters}d. In \S\ref{sec:stripe_param_opt}, we present the optimisation of two tunable parameter of \texttt{STRIPE}. 

%% file: sections/results.tex
\section{Results and benchmarking}\label{sec:results}
In this section, we present the results from SPOTLIGHT's two-stage RFI filters. Using a large sample of pulsar data, we optimise the \texttt{STRIPE} parameters in \S\ref{sec:stripe_param_opt} and, using the best parameter, compare \texttt{STRIPE} with PRESTO's (\citep{ransom_presto_2011}) RFI mitigation tool in \S\ref{sec:stripe_vs_rfifind}. In \S\ref{sec:single_pulse}, we present the results of RFI filtering for single-pulse recovery, and lastly in \S~\ref{sec:trigger_rate} we discuss the effect of RFI filters on the real-time trigger rate within the SPOTLIGHT system.

\subsection{Optimisation of \texttt{STRIPE} Parameters}\label{sec:stripe_param_opt}

The performance of \texttt{STRIPE} was optimised using a representative sample of 70 beamformed pulsar observations commensally recorded by the SPOTLIGHT system over approximately one year (April 2025 to April 2026). The dataset spans all three SPOTLIGHT observing bands (300–1460 MHz) and samples a wide range of observing conditions encountered during routine telescope operations. All observations were recorded in the PC beamforming mode, using the standard SPOTLIGHT configuration with 4096 frequency channels and a sampling time of 1.31072 ms. Table~\ref{tab:pulsar_props} presents the physical properties of these pulsars, spanning a wide variety of periods, flux densities, dispersion measures, and duty cycles, making the dataset representative of the diverse observing conditions under which SPOTLIGHT routinely operates. The effectiveness of \texttt{STRIPE} was quantified using the folded signal-to-noise ratio (S/N), as single pulses are not consistently detectable for many pulsars. By combining information from all pulses in an observation, folded S/N provides a robust measure of the cumulative impact of RFI mitigation on pulsar detectability.

Since the voltage-level RFI filter (\texttt{VOLT}) was directly adopted from the GMRT Wideband Backend (GWB), the primary optimisation effort focused on the two tunable parameters within \texttt{STRIPE}: the interquartile range (IQR) threshold ($\kappa$) used by the SK filter and the baseline subtraction window size ($\omega$), as described in \S\ref{sec:stripe}. For each pulsar observation, \texttt{STRIPE} was executed over a combination of values of $\kappa$ and $\omega$. The IQR threshold, $\kappa$, was varied over 25 values between 0 and 30, with finer sampling at low thresholds ($0.0-7$) and coarser sampling at higher thresholds ($8-30$). The window size, $\omega$, was varied from 0.1 to 0.9 in steps of 0.1, resulting in a total of 225 independent parameter combinations per observation. Each realisation was subsequently folded using PRESTO, and the relative change in folded S/N with respect to the original unmitigated data was computed. The parameter combination that yielded the maximum relative change was selected as the optimal configuration for that observation. Repeating this procedure for all 70 observations resulted in 15750 (225 $\times$ 70) independent realisations, providing a statistically meaningful basis for parameter optimisation. The mean of the optimal IQR threshold and width were 3.986 and 0.513, respectively. Based on this, we adopted ensemble mean values ($\kappa=4.0$) and ($\omega=0.5$) as the default operating parameters for \texttt{STRIPE} within the SPOTLIGHT pipeline. 

Although these values are not necessarily optimal for every individual observation, they consistently yield near-optimal performance across the entire dataset without requiring dynamic parameter selection. It is important to note that these parameters have been optimised specifically for SPOTLIGHT's observing configuration, frequency coverage, and RFI environment, and should not be regarded as universally optimal for other telescopes or transient-search pipelines. The optimised parameter set is used throughout the remainder of this work, including benchmarking against PRESTO's \texttt{rfifind}, as presented in \S\ref{sec:stripe_vs_rfifind}.

\subsection{\texttt{STRIPE} vs \texttt{rfifind}}\label{sec:stripe_vs_rfifind}
We benchmarked \texttt{STRIPE} against PRESTO's \texttt{rfifind}, one of the most widely used offline RFI mitigation tools in pulsar astronomy. Although \texttt{rfifind} is primarily intended for offline processing, it provides a useful reference for evaluating the effectiveness of RFI mitigation. The comparison was performed using the same pulsar dataset described in \S\ref{sec:stripe_param_opt}, ensuring that both parameter optimisation and performance evaluation were based on a common observational sample. We excluded $\sim$21\% of the observations in which \texttt{rfifind} corrupted the original input data. The effectiveness of both methods was evaluated using the folded S/N obtained from PRESTO. Both \texttt{STRIPE} and \texttt{rfifind} were executed with a block size of 1526 samples ($\sim$2 seconds), using 8 CPU cores for RFI mitigation and 16 CPU cores for folding. \texttt{STRIPE} was run with the default operating parameters adopted for SPOTLIGHT ($\kappa=4.0$ and $\omega=0.5$), while \texttt{rfifind} was executed using its standard configuration, including 10$\sigma$ time-domain and 4$\sigma$ frequency-domain rejection thresholds, and 6$\sigma$ time-domain clipping.

\begin{figure}
    \centering
    \includegraphics[width=1\linewidth]{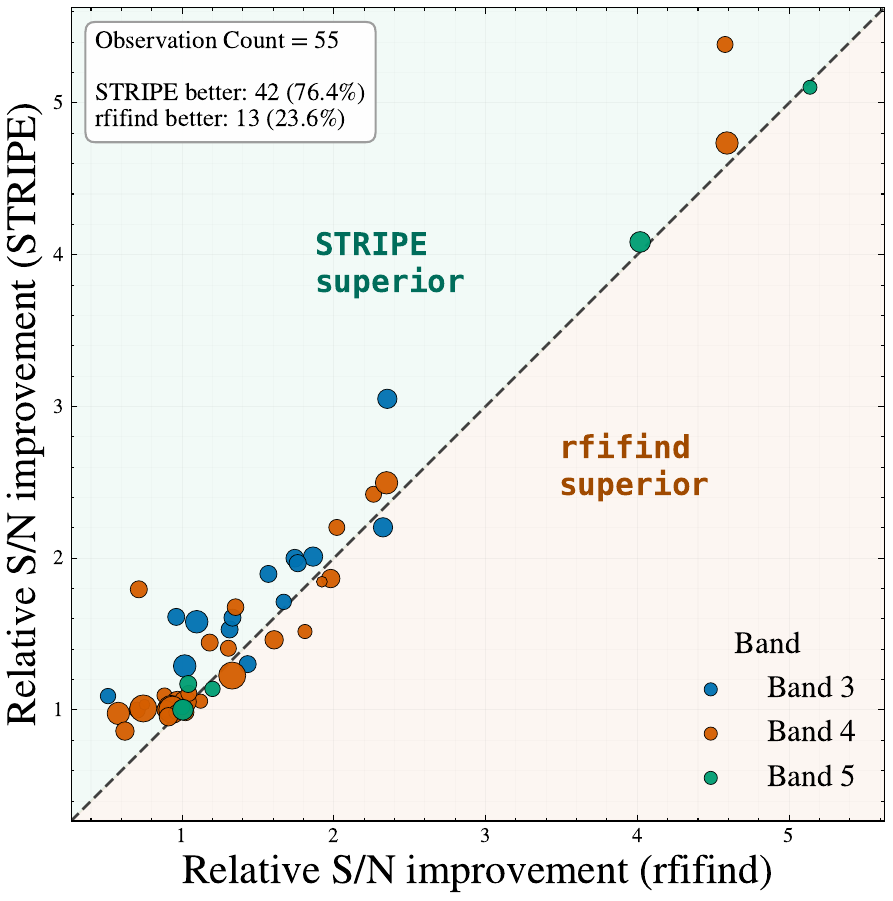}
    \caption{Relative folded S/N improvement with \texttt{STRIPE} versus \texttt{rfifind} for 55 pulsar observations. Points are colour-coded by observing band and scaled by pulsar duty cycle; the dashed $y=x$ line denotes equal performance, with points above (below) the line favouring \texttt{STRIPE} (\texttt{rfifind}).}
    \label{fig:stripe_rfifind_snr}
\end{figure}

Figure~\ref{fig:stripe_rfifind_snr} compares the relative S/N improvement produced by \texttt{STRIPE} and \texttt{rfifind}. The majority of observations ($\sim$76\%) lie above the unity-slope line, indicating greater S/N improvement with \texttt{STRIPE}. In the remaining cases, the difference is generally small, with the measurements clustered close to the equality line. Using the median relative folded S/N improvement as the comparison metric, \texttt{STRIPE} achieved an enhancement that was 17.6\% larger than that obtained with \texttt{rfifind}. Overall, these results indicate that \texttt{STRIPE} provides RFI mitigation performance comparable to \texttt{rfifind}, while yielding greater folded-S/N improvement in a majority of the observations considered here.

\subsection{Single pulse candidate recovery}\label{sec:single_pulse}
\begin{figure*}
    \centering
    \begin{subfigure}{0.8\linewidth}
        \centering
        \includegraphics[width=\linewidth]{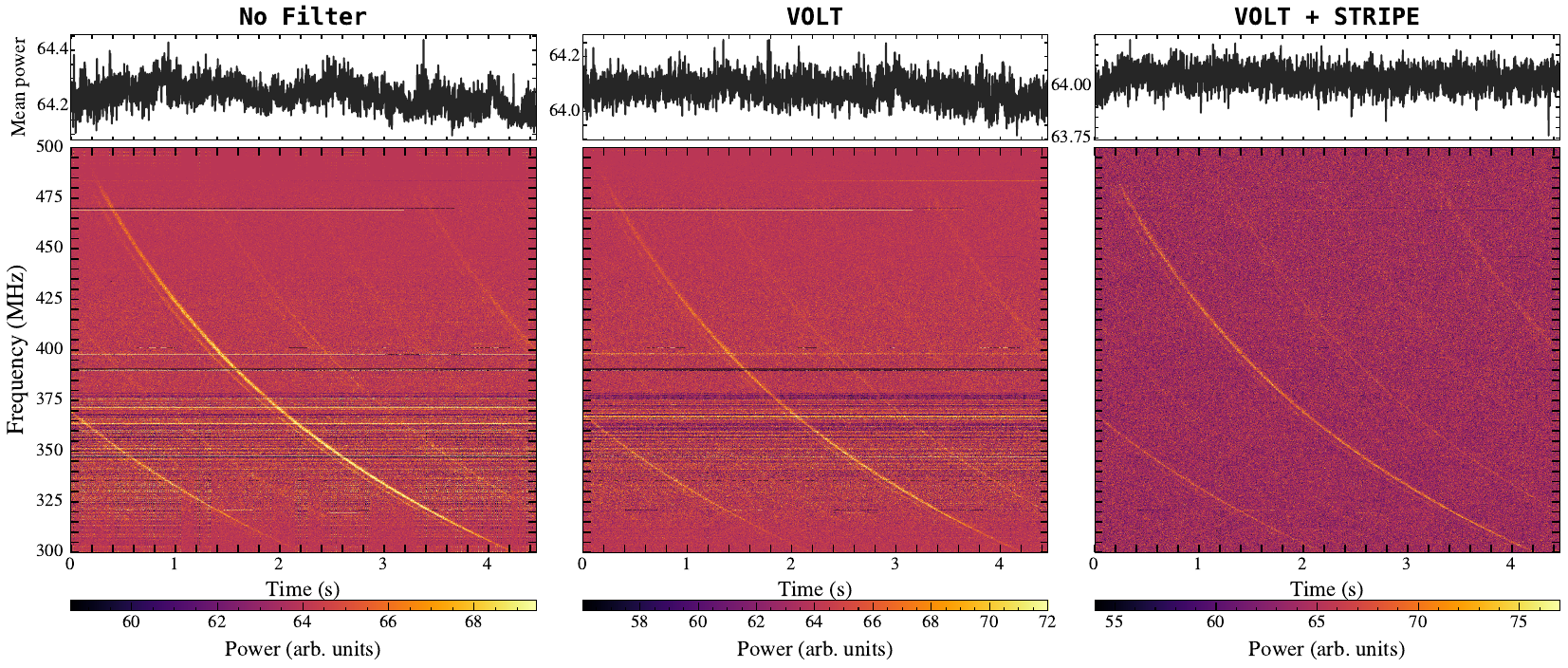}
        \caption{Band 3 (300--500 MHz), PC beam mode, J2113+4644}
        \label{fig:dynspect_comparison_PC_B3}
    \end{subfigure}

    \vspace{0.5em}
    \begin{subfigure}{0.8\linewidth}
        \centering
        \includegraphics[width=\linewidth]{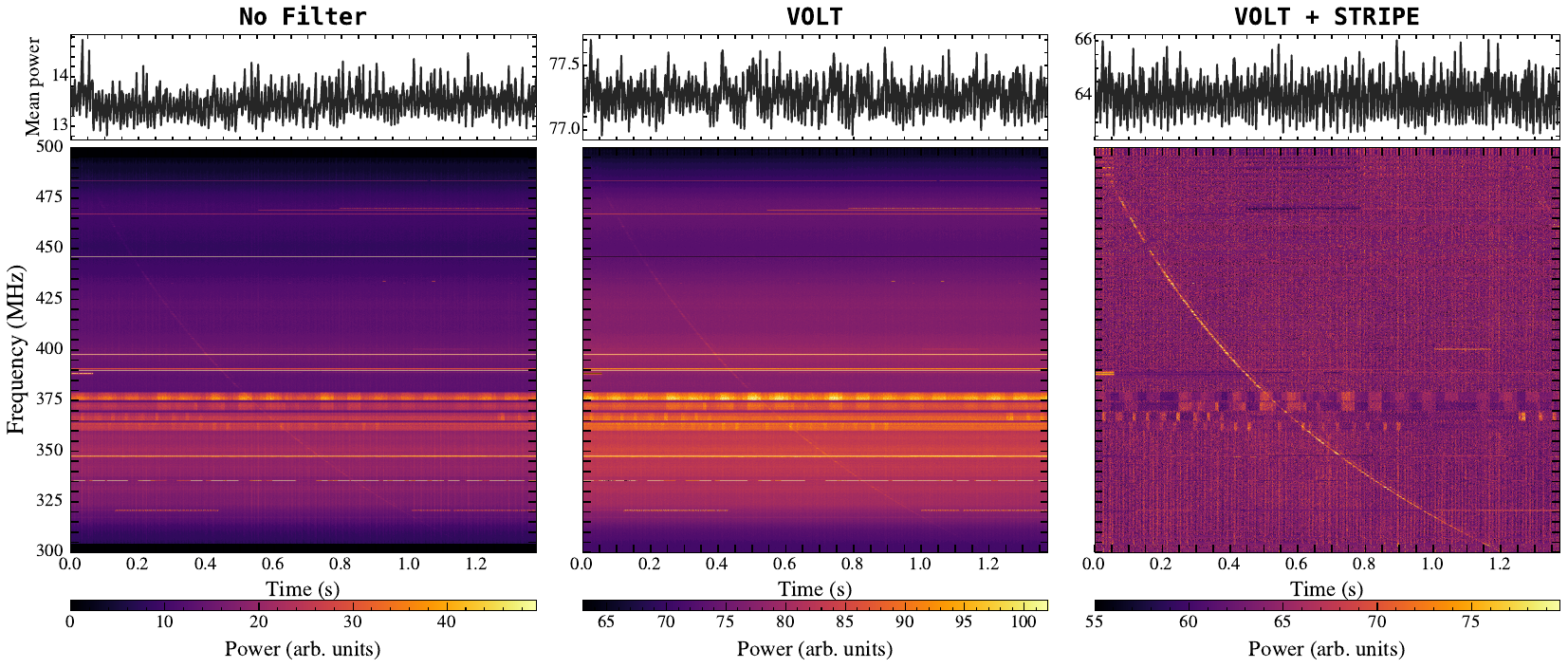}
        \caption{Band 3 (300--500 MHz), IA beam mode, J0820-1350}
        \label{fig:dynspect_comparison_IA_B3}
    \end{subfigure}

    \caption{Comparison of SPOTLIGHT beamformed data processed with
    no RFI filtering, \texttt{VOLT}, and \texttt{VOLT+STRIPE}.
    The dynamic spectra, together with the corresponding time-series
    profiles, show the effect of successive RFI mitigation stages on
    the beamformed data.}
    \label{fig:dynspect_comparison_B3}

\end{figure*}

While the improvement in folded pulsar S/N demonstrates the functional benefit of RFI mitigation, an equally important requirement is that the two-stage filtering preserves, and ideally improves, the detectability of genuine astrophysical transients. To evaluate this, we conducted dedicated observations of six pulsars spanning all three SPOTLIGHT observing bands and compared the recovery of individual pulses under different RFI-mitigation configurations. Figure~\ref{fig:dynspect_comparison_PC_B3} and \ref{fig:dynspect_comparison_IA_B3} illustrate the effect of \texttt{VOLT} and \texttt{STRIPE} on a representative unfiltered beamformed dataset.

Simultaneous observations were carried out using both the GWB and the SPOTLIGHT backend. Since both systems receive the same digitised antenna voltages from the GMRT receiver chain, differences in the recovered pulse populations can be attributed primarily to differences in the RFI-mitigation configurations, while intrinsic source variability and observing conditions are common to the simultaneous observations. To isolate the contribution of each filtering stage, three processing configurations were analysed: (i) GWB data without real-time RFI mitigation, (ii) SPOTLIGHT data processed with \texttt{VOLT} only, and (iii) SPOTLIGHT data processed with both \texttt{VOLT} and \texttt{STRIPE}, with \texttt{STRIPE} applied offline to the recorded beamformed data while preserving the real-time observing configuration.

Beamformed data from each configuration were dedispersed using PRESTO over a narrow dispersion-measure range centred on the catalogue DM of each pulsar\footnote{Pulsar parameters were obtained from the ATNF Pulsar Catalogue: \url{https://www.atnf.csiro.au/research/pulsar/psrcat/}} (DM $\pm 1$ pc cm$^{-3}$, with a step size of 0.1 pc cm$^{-3}$), yielding 20 trial DMs. Single-pulse candidates were subsequently identified using PRESTO's \texttt{single\_pulse\_search.py}. To reduce contamination from residual RFI, only candidates with detection significance above $5\sigma$ and widths below 50 ms (or 100 ms for pulsars with intrinsically broader pulses) were retained. Multiple detections corresponding to the same astrophysical pulse were consolidated by requiring successive candidates to be separated by at least one pulsar rotation period. The remaining candidates were then visually inspected using their dedispersed dynamic spectra and DM transforms. Genuine pulses were required to exhibit a characteristic dispersed track across frequency, along with the expected bow-tie signature in DM space, thereby enabling reliable rejection of residual RFI events. The analysis was performed independently for both PC and IA beam observations across all three SPOTLIGHT observing bands.

\begin{figure}
    \centering
    \includegraphics[width=1\linewidth]{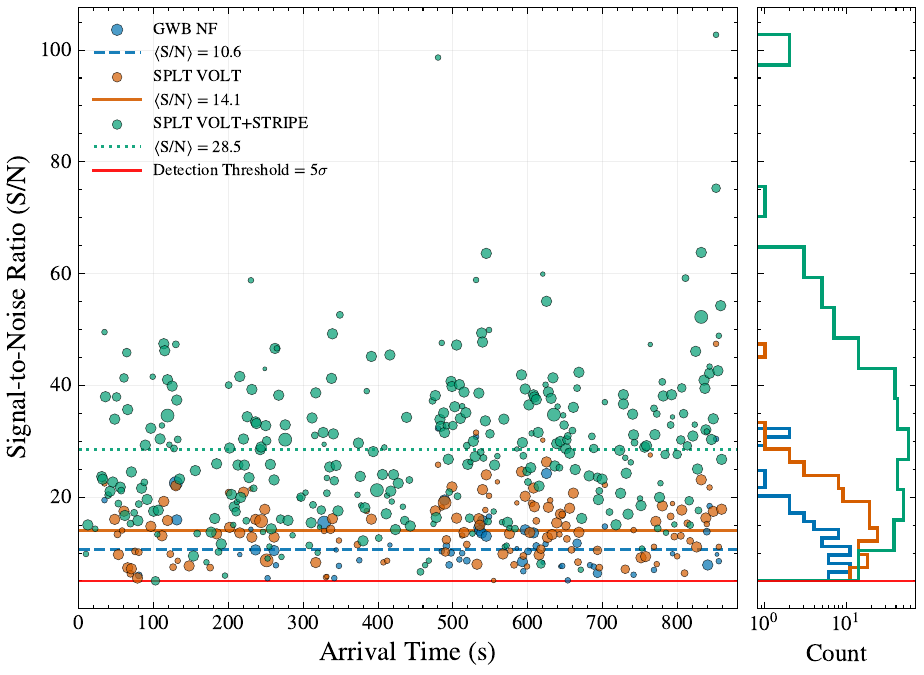}
    \caption{Single-pulse S/N distribution and temporal distribution of detected pulses from J0820$-$1350 in Band 3 (300–500 MHz) for the unfiltered, \texttt{VOLT}-filtered, and \texttt{VOLT}+\texttt{STRIPE}-filtered data. Each point is scaled with detected width, and the horizontal lines in the S/N--arrival-time panel indicate the median S/N for each configuration, while the solid red line marks the $5\sigma$ detection threshold.}
    \label{fig:sps_B3_IA}
\end{figure}

\begin{figure}
    \centering
    \includegraphics[width=1\linewidth]{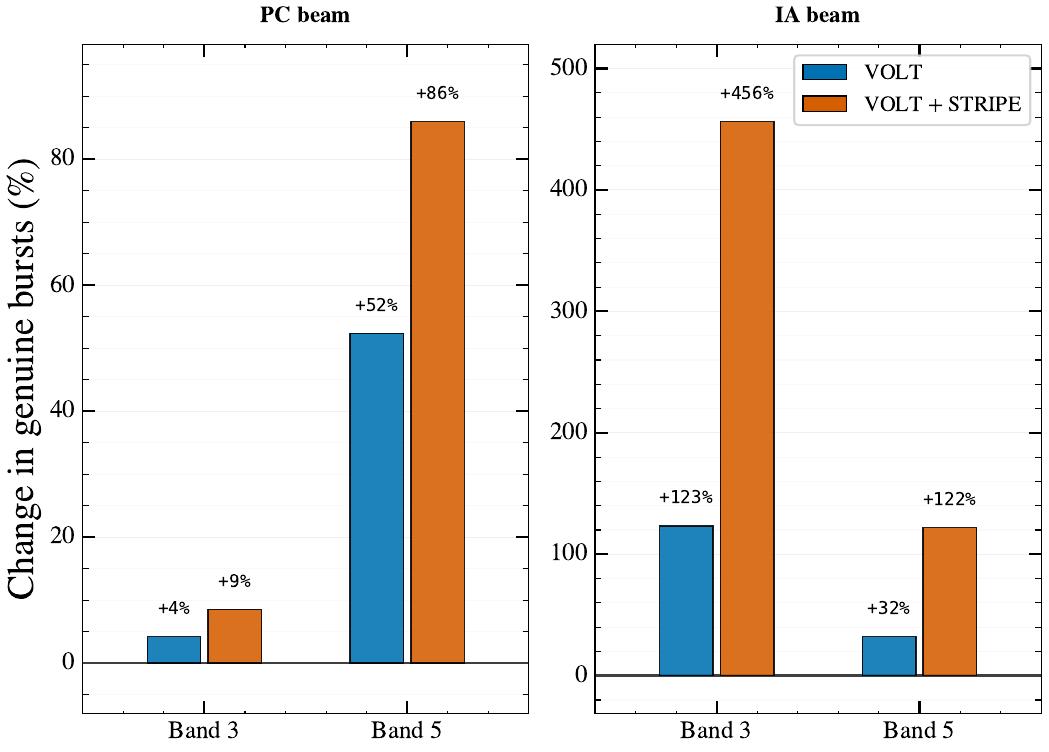}
    \caption{Relative change in the number of genuine single pulses detected after RFI mitigation, with respect to the unfiltered GWB data, for IA and PC beams across Bands 3 and 5.}
    \label{fig:sps_bar_plot}
\end{figure}

The resulting single-pulse populations provide a direct measure of the effect of RFI mitigation on burst recovery. Figure~\ref{fig:sps_B3_IA} shows the S/N of individual pulses as a function of pulse index for a representative Band 3 IA observation. Following the application of \texttt{VOLT}, and subsequently \texttt{VOLT}+\texttt{STRIPE}, additional genuine pulses are recovered, accompanied by an increase in the median single-pulse S/N. The recovered pulses exhibit a 2.7$\times$ improvement in median S/N after the two-stage filtering compared with the unfiltered data. We summarise this behaviour across the complete pulsar sample in Figure~\ref{fig:sps_bar_plot}, which shows the relative change in the number of genuine pulses recovered for the \texttt{VOLT}-only and \texttt{VOLT}+\texttt{STRIPE} configurations for PC and IA observations in Bands 3 and 5. Across the six pulsar observations, the number of genuine pulses increased relative to the unfiltered reference after applying \texttt{VOLT}, with increases of 4\% and 52\% for Band 3 and Band 5 PC beams, respectively, and 123\% and 32\% for the corresponding IA beams. Applying \texttt{VOLT}+\texttt{STRIPE} produced further increases of 9\% and 86\% for Band 3 and Band 5 PC beams, respectively, and 456\% and 122\% for the corresponding IA beams. The increase in the recovered pulse population across different observing configurations indicates that RFI mitigation can recover genuine single pulses not identified in the unfiltered data. For the representative observation shown in Figure~\ref{fig:sps_B3_IA}, the recovered pulses also show an increase in median S/N following RFI mitigation, illustrating the corresponding improvement in single-pulse detectability.

\subsection{Trigger Rate Reduction}{\label{sec:trigger_rate}}
The two-stage RFI mitigation framework was integrated into SPOTLIGHT towards the end of uGMRT observing Cycle 49 (April 2026) and has been in routine operation since the beginning of Cycle 50. The SPOTLIGHT real-time transient detection pipeline is designed to accommodate approximately one trigger every 10 min (Reddy et al. in prep), a limit imposed by the computational and storage resources required for downstream validation, imaging, and archival. For every transient candidate, SPOTLIGHT records high-resolution beamformed data, interferometric raw visibilities, and baseband voltages over the burst duration, together with additional pre- and post-event buffers. For the most demanding observing configuration (Band 3, corresponding to the maximum searched dispersion delay of $\sim$30 s at $DM=1000$ $pc/cm^{3}$), a single trigger results in a maximum data dump of approximately 644 GB, with the baseband (425 GB) and visibility products (219 GB) dominating the storage requirements. At the nominal operating rate of one trigger every 10 min, the available 1 PB storage will be exhausted in only 11.3 days. The derivation of these extreme estimates is provided in Appendix~\ref{appendix:A}. Furthermore, the real-time imaging pipeline (Pal et al. in prep) makes sustained operation at substantially higher trigger rates computationally impractical. Consequently, suppressing false triggers is essential for maintaining uninterrupted commensal observations.

\begin{figure}
    \centering
    \includegraphics[width=1\linewidth]{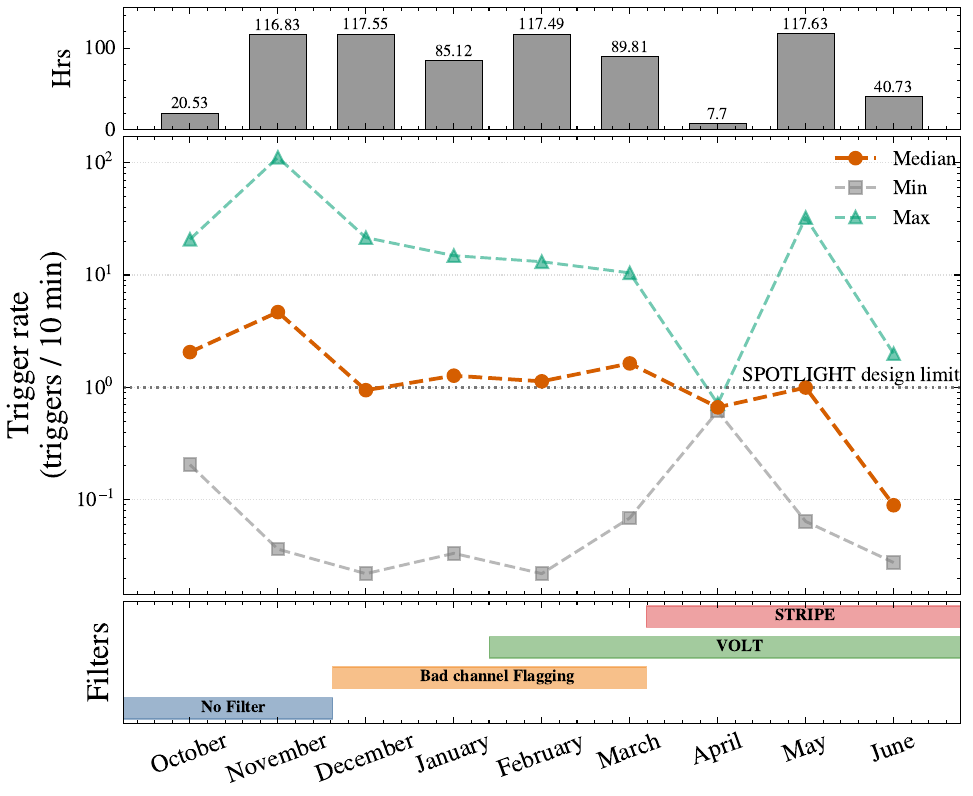}
    \caption{Monthly temporal evolution of the SPOTLIGHT trigger rate from October 2025 to June 2026, expressed as the number of triggers per 10 min. The median, minimum, and maximum trigger rates across observations in each month are shown by separate markers, while the horizontal dashed line indicates the nominal SPOTLIGHT processing limit of one trigger per 10 min. The total observing time contributing to each monthly estimate is shown by the bars in the upper panel; the limited April coverage results in closely clustered statistics.}
    \label{fig:trigger_rate}
\end{figure}

To quantify the operational impact of the proposed RFI mitigation framework, we estimated the trigger rate across all commensally recorded non-pulsar GTAC (GMRT Time Allocation Committee) observations (in total 250 observations spanning more than 713 hours of GTAC time) acquired between the beginning of Cycle 49 (October 2025) and the middle of Cycle 50 (June 2026). Since SPOTLIGHT performs a transient search continuously throughout every observation, irrespective of the primary science target, the trigger rate was computed by normalising the total number of triggers by the full observation duration. Figure~\ref{fig:trigger_rate} presents the monthly median trigger rate per 10 minutes, along with the minimum and maximum trigger rates across observations obtained within each month.

The temporal evolution of the trigger rate closely follows the successive deployment of RFI mitigation strategies in SPOTLIGHT. During October and November 2025, when no real-time RFI mitigation was available, the median trigger rate was approximately 5 triggers per 10 min, which, in an extreme scenario, would completely utilise the available 1 PB of storage in only 2.26 days, well below the timescale required for downstream inspection and data management. In December, a temporary mitigation strategy was introduced by flagging known, persistently RFI-contaminated frequency channels in the beamformed data, resulting in a moderate reduction in trigger rates. Beginning in February, the antenna-level voltage filter (\texttt{VOLT}) was deployed while the bad-channel masking strategy remained in place. Following the deployment of \texttt{STRIPE} in April, the temporary masking strategy was discontinued, and the complete two-stage RFI mitigation framework, comprising \texttt{VOLT} and \texttt{STRIPE}, became operational from the start of Cycle 50. As April 2026 was largely occupied by GMRT maintenance and the transition from cycle 49 to 50, the number of commensal non-pulsar observations was only 2, so the statistics are heavily biased due to the limited sample size.

The deployment of the complete framework reduced the median trigger rate from 5 to approximately 0.09 triggers per 10 min, extending the continuous operating lifetime of the available storage from 2.3 days to 125.5 days, comfortably exceeding the design requirement of one trigger every 10 min. This reduction restores SPOTLIGHT to its intended operating regime while substantially decreasing the burden of false-positive candidates on the transient validation and real-time imaging pipelines. The close correspondence between the deployment timeline and the observed reduction in trigger rate demonstrates that a large fraction of triggers generated during routine commensal observations originated from terrestrial radio-frequency interference. More importantly, the proposed two-stage mitigation framework transforms SPOTLIGHT from a storage-limited system into one capable of sustained, real-time transient searches without compromising its ability to detect genuine astrophysical transients.

%% file: sections/futurework.tex
\section{Future Work}\label{sec:futurework}
Several extensions are planned to further improve the efficiency and adaptability of the framework. In particular, \texttt{STRIPE} can be extended to exploit the spatial information available from the multiple simultaneous beams. Genuine astronomical sources are expected to produce spatially localised signals, whereas terrestrial interference may appear simultaneously across multiple beams. The shared-memory architecture, which provides access to multiple beamformed data streams on each compute node, could therefore be used to construct common RFI masks and reduce redundant channel-level computations across beams. Such spatial information could complement the existing spectro-temporal filtering and multi-beam candidate-selection strategies. A further optimisation is to investigate whether RFI masks derived from the IA beam, which generally experiences a higher RFI burden and requires substantially less computation because only a single beam is processed, can be used to pre-identify persistently contaminated frequency channels in the PC beams. These approaches could reduce the computational cost of \texttt{STRIPE} and provide additional headroom for increasing the number of beams processed in real-time. Future development will also explore adaptive and machine-learning-assisted approaches to improve the robustness of RFI identification under changing interference environments while preserving astrophysical signals.

%% file: sections/summary.tex
\section{Summary}\label{sec:summary}

We have presented a two-stage real-time RFI mitigation framework developed for the SPOTLIGHT transient-search pipeline, comprising the antenna-level voltage filter (\texttt{VOLT}) and the beamformed-data filtering framework (\texttt{STRIPE}). The two filters operate at complementary stages of the signal chain, mitigating impulsive interference before correlation and residual spectro-temporal interference after beamforming. Following their deployment, SPOTLIGHT has experienced a substantial reduction in the trigger rate during routine commensal observations, bringing the system within its intended operational capacity for sustained real-time transient searches.

Tests using pulsar observations demonstrate that the two-stage filtering improves the recovery of genuine single pulses. Comparison with PRESTO's \texttt{rfifind} shows that \texttt{STRIPE} provides comparable or greater improvement in folded S/N for the pulsar sample analysed. Together, these results demonstrate the practical utility of lightweight, real-time RFI mitigation for maintaining the sensitivity and computational efficiency required by high-throughput transient surveys. By mitigating the impact of terrestrial radio-frequency interference, the framework enhances SPOTLIGHT’s sensitivity to weaker radio transients, potentially improving the detectability of faint pulsars and FRBs.

%% file: sections/thanks.tex
\begin{acknowledgments}
We acknowledge constructive discussions with Yunpeng Men and Emma Carli, which helped us better understand the RFI mitigation algorithm. We acknowledge the invaluable contribution of colleagues from the Oxford e-Research Centre (OeRC), NVIDIA, and the Centre for the Development of Advanced Computing (C-DAC). We thank our colleagues at the Centre for Development of Advanced Computing (C-DAC) for their support in setting up the Param Brahmand data centre at the GMRT. We acknowledge funding for the SPOTLIGHT backend (called Param Brahmand) under the National Supercomputing Mission (NSM) Phase 3, as well as support from the Department of Atomic Energy (DAE), Government of India, under project no. $\rm 12-R\&D-TFR5.02-0700$, for the contributions towards the overhead cost. The uGMRT is operated by the National Centre for Radio Astrophysics of the Tata Institute of Fundamental Research, India. We also acknowledge the DAE for the fund under the project titled Next Generation Instrumentation for Radio Astronomy (project UID number being RTI4017).

We gratefully acknowledge support from the ``Building Indo–UK Collaborations Towards the Square Kilometre Array'', funded under the DAE–STFC Technology and Skills Programme, which facilitated the development of SPOTLIGHT's highly efficient, real-time transient search pipelines. We sincerely thank the uGMRT engineers involved in the SPOTLIGHT survey for their relentless efforts in commissioning and stabilising SPOTLIGHT's correlator and beamformer systems, ensuring high-quality data flow for all of SPOTLIGHT's real-time and offline systems. We also thank the uGMRT operators for their coordinated efforts in successfully conducting the SPOTLIGHT survey and test observations. This research was supported in part by the International Centre for Theoretical Sciences (ICTS) for the FTSky: A program in the field of Fast Radio Transients (code: ICTS/FTSky2025/10). The PI gratefully acknowledges the financial support provided by Premji Invest’s generous donation to TIFR to pursue the application of AI to accelerate scientific discovery in astrophysics. This funding helped support a Research Assistant position at NCRA–TIFR and enabled this research.

\end{acknowledgments}

%% file: sections/appendix.tex
\appendix
\section{Pulsar Data}
Table~\ref{tab:pulsar_props} presents the physical properties of pulsars used in the optimisation, comparison of \texttt{STRIPE} and \texttt{rfifind}, as well as single pulse recovery. The dataset spans a wide range of periods, flux densities, dispersion measures, and duty cycles, making it representative of the diverse observing conditions under which SPOTLIGHT routinely operates.

\begin{table}
\centering
\caption{Physical properties of the pulsars used throughout this paper to evaluate the performance of the RFI mitigation algorithms. Dispersion measures, periods, pulse widths, duty cycles, and flux densities (at 1400 MHz) are taken from the ATNF Pulsar Catalogue \citep{manchester_australia_2005}.}
\label{tab:pulsar_props}
\begin{tabular}{lrrrrr}
\toprule
Pulsar & DM (pc cm$^{-3}$) & Period (s) & $W_{50}$ (ms) & Duty Cycle (\%) & Flux Density (mJy) \\
\midrule
J0139$+$5814 & 73.81 & 0.27 & 5.20 & 1.90 & 4.60 \\
J0452$-$1759 & 39.90 & 0.55 & 27.00 & 4.91 & 17.00 \\
J0534$+$2200 & 56.77 & 0.03 & 2.00 & 6.00 & 14.00 \\
J0543$+$2329 & 77.70 & 0.25 & 4.70 & 1.90 & 10.70 \\
J0630$-$2834 & 34.42 & 1.24 & 63.00 & 5.06 & 32.00 \\
J0659$+$1414 & 13.95 & 0.39 & 14.80 & 3.80 & 2.70 \\
J0729$-$1448 & 91.93 & 0.25 & 7.60 & 3.00 & 0.83 \\
J0729$-$1836 & 61.22 & 0.51 & 14.00 & 2.70 & 1.90 \\
J0742$-$2822 & 73.76 & 0.17 & 4.40 & 2.60 & 26.00 \\
J0820$-$1350 & 40.94 & 1.24 & 23.00 & 1.86 & 6.00 \\
J0835$-$4510 & 67.77 & 0.09 & 1.70 & 1.90 & 1050.00 \\
J1644$-$4559 & 478.66 & 0.46 & 8.00 & 1.76 & 300.00 \\
J1705$-$3423 & 146.15 & 0.26 & 12.00 & 4.70 & 5.30 \\
J1709$-$4429 & 75.58 & 0.10 & 6.00 & 5.90 & 12.10 \\
J1731$-$4744 & 123.06 & 0.83 & 18.00 & 2.20 & 27.00 \\
J1740$-$3015 & 151.90 & 0.61 & 3.00 & 0.50 & 8.90 \\
J1751$-$3323 & 295.54 & 0.55 & 7.50 & 1.40 & 1.67 \\
J1752$-$2806 & 50.323 & 0.56 & 6.60 & 1.17 & 48.00 \\
J1803$-$2137 & 234.04 & 0.13 & 13.00 & 9.70 & 9.60 \\
J1825$-$0935 & 19.38 & 0.77 & 11.00 & 1.40 & 10.00 \\
J1847$-$0402 & 141.98 & 0.60 & 20.00 & 3.30 & 4.90 \\
J1932$+$1059 & 3.18 & 0.23 & 5.60 & 2.50 & 29.00 \\
J1935$+$1616 & 158.64 & 0.36 & 6.50 & 1.81 & 58.00 \\
J2113$+$4644 & 141.26 & 1.01 & 60.50 & 5.96 & 19.00 \\
\bottomrule
\end{tabular}
\end{table}

\section{Estimation of Maximum Trigger Data Volume}\label{appendix:A}
The maximum data volume generated by a single SPOTLIGHT trigger is determined by the longest possible dispersive delay searched by the real-time transient pipeline. This appendix derives the trigger size quoted in \S\ref{sec:trigger_rate}, which is subsequently used to estimate the operational storage lifetime of the SPOTLIGHT system.

Throughout routine observations, SPOTLIGHT records three independent data products for every detected transient candidate: (i) beamformed data, (ii) high-time-resolution interferometric visibilities, and (iii) raw baseband voltages. The amount of data recorded depends on the burst duration, the searched dispersion measure (DM), and the buffering strategy adopted by the pipeline. The largest data volume occurs for the lowest observing frequency and the highest searched dispersion measure, since these produce the maximum dispersive delay across the observing band.

For the standard Band 3 transient search, SPOTLIGHT searches up to a maximum dispersion measure of ($DM_{max}$) of 1000 $pc/cm^3$, which corresponds to a maximum dispersive delay ($\Delta t_{DM}$) of 29.50 seconds (\cite{kulkarni_dispersion_2020}). The following estimates assume a negligible intrinsic burst width so that the data volume is dominated by the dispersion delay.

\subsection{Beamformed Data}
The dumped beamformed data consists of the complete dispersed burst together with equal pre- and post-event margins. Consequently, the number of beamformed samples written to disk is
\begin{align*}
    N_{beam}= 2\times N_{delay} + N_{burst} \approx 2 N_{delay}
\end{align*}
where $N_{delay}=45017$ samples for the maximum dispersion delay at the default sampling interval of $t_{samp}=1.31072$ms. This corresponds to a beamformed data volume of $S_{beam}  \approx 175.8 \text{ MB}$.

Since beamformed data are total-intensity products, they contribute only a small fraction of the total trigger volume.

\subsection{Visibility Data}
The real-time imaging pipeline requires high-time-resolution interferometric visibilities surrounding each transient candidate. To accommodate pipeline latency, the visibility dump can include at most one complete buffer before and after the dispersed burst and an additional 10\% timing margin on either side of the dispersion delay. For the current SPOTLIGHT correlator,
\begin{align*}
    N_{vis} = 2 \times \binom{32}{2} + 64=1056
\end{align*}
visibility products are recorded, comprising all cross-correlations for two polarisations together with the autocorrelation products from 32 antennas. Each complex visibility is stored as a half-precision floating-point number (2 bytes per component). The resulting visibility dump occupies $S_{vis}  \approx 219.1 \text{ GB}$.

\subsection{Baseband Data}
The largest contribution to the trigger volume arises from the raw baseband voltages. Similar to the visibility dump, the baseband recording includes the complete dispersed burst, along with at most one buffer before and after the event, and an additional 10\% timing margin. The SPOTLIGHT correlator records voltages from 32 antennas $\times$ 2 polarisations = 64 independent baseband streams. For the standard 200 MHz observing mode, the voltages are represented using 4-bit samples. Each baseband buffer contains 400 MB, corresponding to
1.048576 seconds of data. Accounting for the complete dispersed burst and the required buffering yields a total baseband data volume of $S_{baseband} \approx 425 \text{ GB}$.

\subsection{Total Trigger Size}
The maximum data volume associated with a single transient trigger is therefore
\begin{align*}
    S_{trigger} &= S_{beam} + S_{vis} + S_{baseband} \\
    &= 175.8 \text{ MB} + 219.1 \text{ GB} + 425 \text{ GB} \\
    &\approx 644.3 \text{ GB}
\end{align*}
The beamformed data contribute less than 0.1\% of the total trigger size, while the visibility and baseband products dominate the storage requirements.

Assuming an effective storage capacity of approximately 1 PB, the maximum number of triggers that can be retained simultaneously is
\begin{align*}
    N_{trigger} = \frac{1 \text{ PB}}{644.3 \text{ GB}} \approx 1627 \text{ triggers}
\end{align*}
At the nominal design rate of one trigger every 10 min, the available storage would therefore be exhausted after 1627$\times$10 min = 11.3 days. 

In contrast, as described in \S\ref{sec:trigger_rate}, the median trigger rate observed prior to the deployment of the real-time RFI mitigation framework was approximately 5 triggers per 10 min, reducing the effective storage lifetime to only 2.3 days. Following the deployment of \texttt{VOLT} and \texttt{STRIPE}, the median trigger rate decreased to 0.09 triggers per 10 min, extending the storage lifetime to approximately 125.5 days. This increase in operational lifetime illustrates the practical importance of real-time RFI mitigation in enabling sustained commensal observations with SPOTLIGHT.